\documentclass{PoS}

\usepackage[square]{natbib}
\title{Radio Galaxy Zoo: host galaxies and radio morphologies for large surveys from visual inspection}

\ShortTitle{Radio Galaxy Zoo}

\author{\speaker{Kyle W. Willett}\thanks{On behalf of the Radio Galaxy Zoo science team.}\\
        School of Physics and Astronomy, University of Minnesota, 116 Church St. SE, Minneapolis, MN 55455, USA\\
        E-mail: \email{willett@physics.umn.edu}}

\abstract{We present a description and early results from Radio Galaxy Zoo, a web-based citizen science project for visual inspection and classification of all-sky radio surveys. The goals of the project are to classify individual radio sources (particularly galaxies with multiple lobes and/or complex morphologies) as well as matching the continuum radio emission to the host galaxy. Radio images come from the FIRST and ATLAS surveys, while matches to potential hosts are performed with infrared imaging from WISE and SWIRE. The first twelve months of classification have yielded more than 1~million classifications of more than 60,000 sources. For images with at least $75\%$ consensus by the volunteer classifiers, the accuracy is comparable to visual inspection by the expert science team. Based on mid-infrared colors, the hosts associated with radio emission are primarily a mixture of elliptical galaxies, QSOs, and LIRGs, which is in good agreement with previous studies. The full catalog of radio lobes and their host galaxies will measure the relative populations of host types as a function of radio morphology and power. Radio Galaxy Zoo has also been an effective method for detecting extremely rare objects, such as HyMORs and giant radio galaxies. Results from the project are being used to train automatic algorithms for host matching for use in future large continuum surveys such as EMU, as well as establishing roles for citizen science in projects such as the SKA.}

\FullConference{EXTRA-RADSUR2015 (*)\\
		20--23 October 2015\\
		Bologna, Italy

                \bigskip
                \hrule
                \bigskip

                \textnormal{(*) This conference has been organized
                  with the support of the Ministry of Foreign Affairs
                  and International Cooperation, Directorate General
                  for the Country Promotion (Bilateral Grant Agreement
                  ZA14GR02 - Mapping the Universe on the Pathway to
                  SKA)}
}

\begin{document}

\section{Motivation and science}

One of the most powerful methods of exploring the evolution of structure in the Universe is through the advent of large-scale radio surveys. Galaxies emit radio-wavelength radiation through a variety of physical mechanisms, both thermal and non-thermal. Radio-wavelength emission is also associated with a wide range of physical processes, including shocks and interactions in the intracluster medium for galaxy clusters, accretion and jets for galaxies with central supermassive black holes, and distributed emission co-located with star formation. 

To transform radio emission into physical measurements of individual galaxies, the radio emission must have a definitive association with a galaxy as inferred from either stellar or quasar-like emission, typically in optical or infrared wavelengths. This step is critical since the redshift of a galaxy cannot be determined from radio continuum emission; measurement of the distance (and thus a transformation from reference-dependent values, such as flux density or angular size, to reference-independent values such as power or linear size) enables the majority of the relevant astrophysics. 

\begin{figure*}
\centering
\includegraphics[angle=0,width=5.0in]{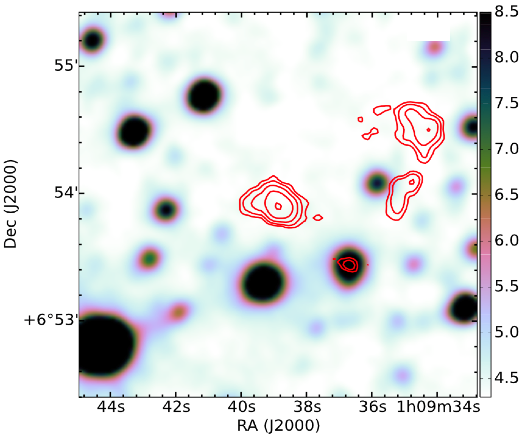}
\caption{Example of a radio galaxy for which simple positional source matching fails. Red contours show the radio emission from the FIRST survey, while the background heatmap is 3.4~$\mu$m emission from the WISE survey. In this case, the high density of IR sources makes spurious associations likely when simply matching radio components to their nearest counterparts. However, even an untrained/lightly-trained human can identify four large radio components as a single galaxy, with its IR counterpart coincident with the bottommost radio component. (Original image from \cite{ban15}, Figure 5e.)
\label{fig-example}}
\end{figure*}

Simple positional matching of radio continuum to optical/IR data of the same field can miss significant numbers of sources in the field. Radio emission, especially from jets associated with black holes, can have a very large angular size as seen projected on the sky that extends well beyond the optical extent of the galaxy. Furthermore, jets are often resolved into multiple discrete components which can be arranged in either linear or bent fashion. There is also no guarantee that a radio source will have core emission overlapping with the galaxy itself. So while automated matching of sources based on position will identify many potential radio-loud galaxies, many will either be mis-identified or omit extended components (see Figure~\ref{fig-example}). 

Visual inspection of radio images overlaid with optical/IR improves on simple positional matching, and to date has been used successfully with data from many surveys. However, the amount of data capable of being processed will scale far beyond the ability of professional astronomers given the expected outputs of surveys such as EMU \citep{nor11}, with more than $10^7$ detections of extragalactic radio sources expected. Associating radio emission with its host can be done without extensive astrophysical training and is not limited to the professional community alone. We present the design and early output of Radio Galaxy Zoo (RGZ), a web-based citizen science project \cite{ban15} that is matching radio emission to near-IR hosts for $\sim175,000$~galaxies from the FIRST \cite{whi97} and ATLAS \cite{fra15} surveys. 

\section{Current methods for morphologically classifying radio emission}

Several automated methods (eg, nearest neighbor) exist to model radio morphologies with their optical/IR counterparts. For example, linear triple systems in the ATLAS survey are modeled by assessing whether multi-component sources are well-fit by variations on lobe-core-lobe morphologies \cite{fan15}. This method uses a Bayesian prior to constrain the likely separation between components as well as the apparent bending angle. A likelihood ratio is used to identify the most probable arrangement, which is then compared to the result of expert visual classification \cite{nor06} to measure the accuracy. The algorithm has a fairly high success rate, but mis-characterizes $\sim10\%$ of triple sources and $\sim30\%$ of double sources. It is also limited by being trained for specific geometries; rare (but extant) systems either with many components or displaying non-linear structures require additional training.

Other automated algorithms are similarly limited to particular systems, typically double sources with a single optical/IR component \cite{pro06,van15}. Complex radio morphologies require more computational resources and are limited by the amount of realistic training data available. 

Expert visual classification can also be limited by high densities of background galaxies and non-linear sources. However, the possibilities for complex morphologies extend to dozens of potential categories, including rare arrangements such as X-shaped, ringed, narrow- or wide-angled tails, etc. \cite{pro11}. The primary impediment to this is scale; an individual or small team can do this for hundreds or thousands of sources \citep{nor06}, but becomes impractical for the surveys with $>10^4$ sources. 

\section{RGZ interface}

\begin{figure*}
\includegraphics[angle=0,width=6.0in]{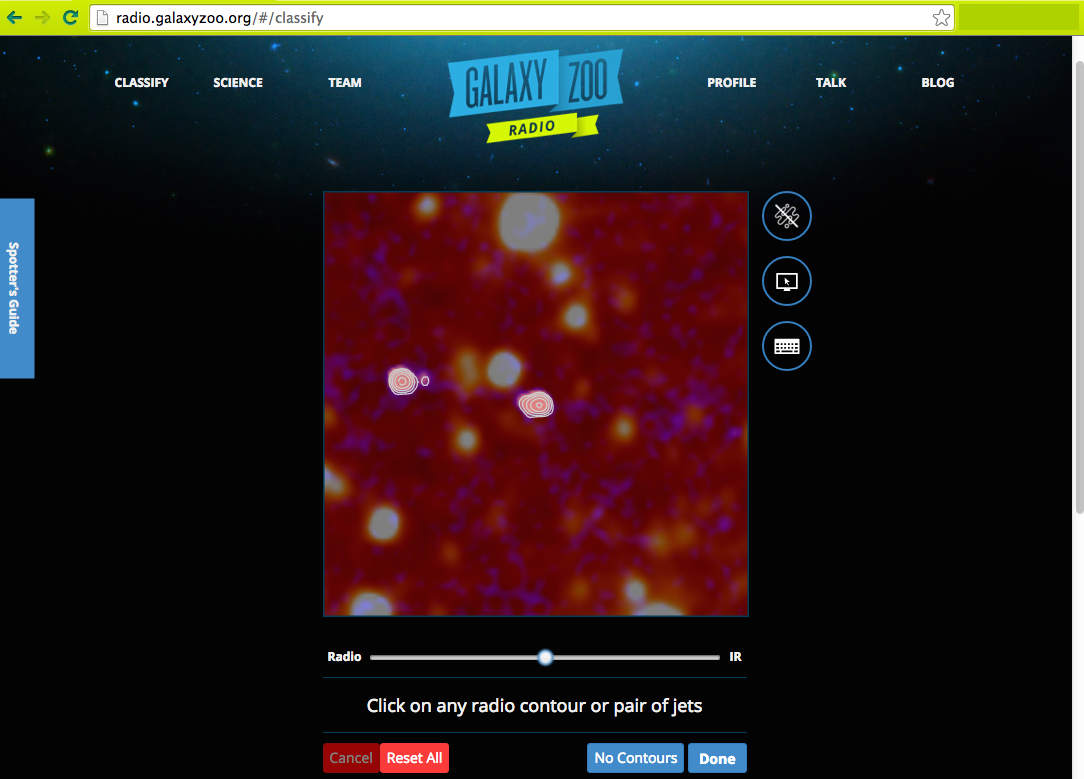}
\caption{Screenshot of a randomly selected FIRST subject in the Radio Galaxy Zoo interface. The slider at the bottom controls the relative opacity between the overlaid radio and infrared images. Users select discrete radio component(s) and identify the location of the infrared counterpart(s), if present.
\label{fig-interface}}
\end{figure*}

The RGZ project is operated by the Zooniverse, the world's largest citizen science platform and host to several dozen citizen science projects. In contrast to the original Galaxy Zoo project \cite{lin08,wil13}, in which users applied discrete labels from a hierarchical decision tree, RGZ classification has a two-step process with direct annotations on the image itself. 

When classifying on the website\footnote{\texttt{http://radio.galaxyzoo.org}}, users are presented with a single panel in which radio data is overlaid on an infrared background (Figure~\ref{fig-interface}). The radio data are pre-processed to identify only strong components, using a median absolute deviation statistic to select objects with flux density peaks above $3\sigma$. These are outlined as discrete components that the user can select by clicking on them; the full radio image is present as a blue-tinted layer, the opacity of which can be controlled by the user as they desire. The standard image size shown to a user is $3\times3$~arcmin. 

In a single classification, users select all radio emission they consider to be originating from a single radio galaxy by clicking on individual components in the image. After selecting the radio components, they advance one step and then click on the image to identify the probable source of the host galaxy based on the near-IR data. Users also have the option of selecting ``No IR counterpart''. If there are multiple radio sources in the image, users can repeat both steps to identify other radio galaxies. Once the user clicks `Done', the classifications are recorded and a new, randomly-selected image appears. Users always have the option to discuss the data with other volunteers and scientists in a discussion forum dubbed `Radio Talk'\footnote{\texttt{http://radiotalk.galaxyzoo.org}}. 

There are two sets of radio/IR data being processed in the initial phase of RGZ. The first selects radio sources in the FIRST catalog \cite{whi97} taken with the Very Large Array, which constitutes a subset of 174,821 resolved sources. The near-IR images matched to host galaxies come from the 3.4~$\mu$m (W1) band from the WISE survey \cite{wri10}. The second set of 4,396~images focuses on a smaller but deeper field centered on ELAIS~S1 and the CDFS. Radio data come from the ATLAS survey \cite{fra15} taken with the Australia Telescope Compact Array and 3.6~$\mu$m near-IR images in the SWIRE survey \cite{lon03a} from the \textit{Spitzer Space Telescope}. 

\section{RGZ data products}

An image is retired from the system once a minimum number of individuals have classified it; that threshold is set to 5~users for images with only a single radio component (in this case, the only ambiguity is the location of the potential IR host) and at 20~users for images with multiple radio components. Individual classifications are aggregated to provide a consensus classification of the image based on the majority vote. This is done by first taking the most common arrangement of the discrete radio sources; using only those classifications, a kernel density estimator is then applied to estimate the location of the IR counterpart. This is repeated if the plurality answer identified more than one radio galaxy within an image.

Classifications for RGZ are stored in a Mongo database, with separate annotations linked to discrete radio components and the $(x,y)$ locations in which users identified the IR hosts. Information from the original WCS header on the FITS cutouts is then used to translate the annotations into J2000 positions. The position of the host galaxy is then cross-matched to both the AllWISE catalog, which locates the nearest object in the W1 image within 3~arcsec, and the SDSS DR12 \texttt{Galaxy} table, which locates an optical counterpart. If either catalog gives a successful cross-match, photometric and spectroscopic information about the likely host are added as metadata to the radio source in our catalog. 

Radio data from the catalog include measurements of angular size and position angle, number of components, peak and integrated flux densities. If a redshift of the galaxy is known, we add data such as distance, physical size, and peak and integrated luminosities. 

\section{Early science results}

The project design and some early science results have been published in \cite{ban15}. Below we describe three projects in progress using RGZ data.

\subsection{Infrared colors of powerful radio galaxy hosts}

\begin{figure*}
\centering
\includegraphics[angle=0,width=5.0in]{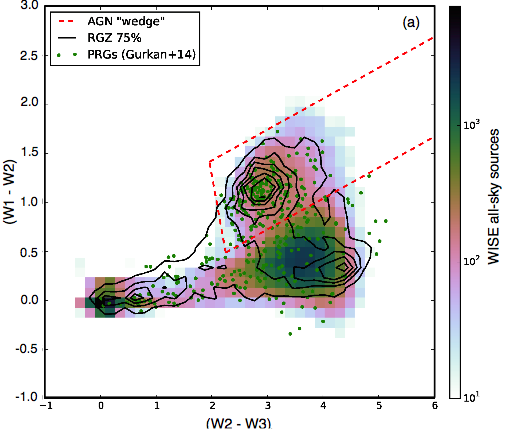}
\caption{Plot of $\sim33,000$ radio hosts identified from RGZ (black contours), overlaid on a subset of the all-sky sources from the WISE catalog (colorscale). Green dots show a smaller comparison sample of radio-loud galaxies \cite{gur14}, while the dashed red line indicates the location of an ``AGN wedge'' identified through X-ray and radio cross-matching. (Adapted from Figure~10 in \cite{ban15}.)
\label{fig-wisecolorcolor}}
\end{figure*}

Since users in the RGZ project are matching the radio emission to near-IR images, this allows for a very large study of the global properties of the host properties as a function of their IR photometry. Using a preliminary set of 33,127 sources classified by RGZ in its first 12~months, we examined the (W1-W2) and (W2-W3) colors of galaxies with a reliable radio association and detection in the WISE all-sky catalog (Figure~\ref{fig-wisecolorcolor}). We find that the majority of radio hosts have WISE colors consistent with elliptical galaxies, QSOs, or LIRGs. There is also a significant population of galaxies with redder colors $[(W2-W3) > 0.5]$ than normal ellipticals; this could be due either to an association of radio emission with star-formation, or to enhanced dust content (Willett et al., in prep). 

\subsection{Radio galaxies with hybrid morphology}

Hybrid radio galaxies (HyMORs) are rare examples of radio galaxies which show differing Fanaroff-Riley (FR) morphologies on either side of an active nucleus \citep{gop00}. Since the difference between the host populations of FRI/FRII galaxies have important implications on the effects of nature vs. nurture hypotheses for radio-loud galaxies, HyMORs are unique test cases that can help to test different physical scenarios. However, such galaxies are extremely rare, with fewer than $50$~candidates reported in the literature. 

Using RGZ data tagged in the Radio~Talk interface, we have been able to identify 11~new HyMORs and two additional candidates based on FIRST imaging. The increased sample size has also allowed more stringent definitions of how to classify a HyMOR and revises the classifications of several previous candidates (Kapi{\'n}ska et al, submitted). 

\subsection{Jet bending angles}

Galaxies with radio jets can have their apparent position angle affected by a number of physical interactions; one such effect arises from the peculiar motion of the galaxy with respect to a dense intracluster medium (ICM) due to ram pressure. In simulations, higher relative velocities are shown to result in narrower tails, which can depend (on average) on the distance to the center of the cluster. Measuring the jet bending angle can yield a powerful probe of the ICM pressure, dynamical state, and merger history.

\begin{figure*}
\centering
\includegraphics[angle=0,width=5.0in]{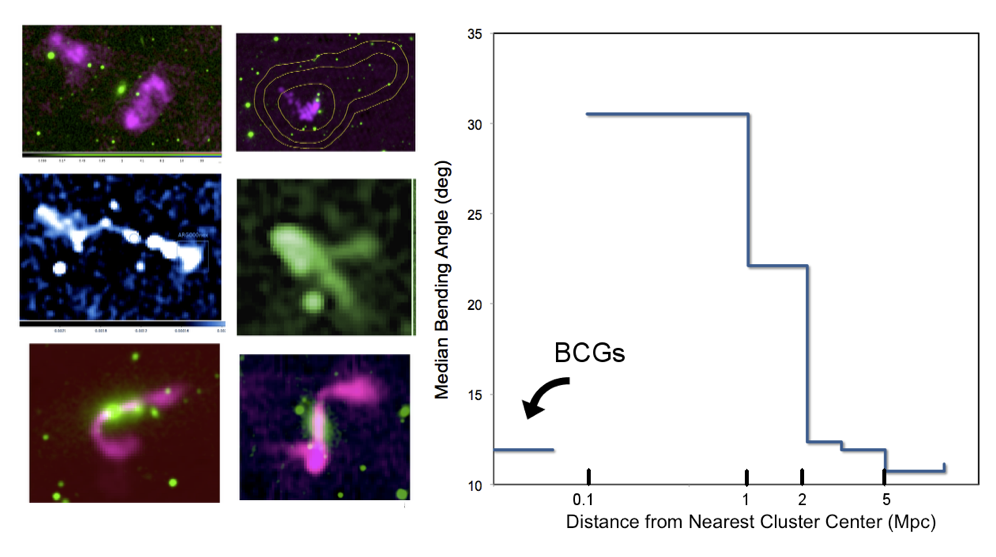}
\caption{\textit{Left}: examples of sharply-bent radio galaxies identified in RGZ. Images are FIRST and NVSS images, zoomed out to show the full scale of the jets for each galaxy. \textit{Right}: Binned plot of the median bending angle from 888~consensus double and/or triple sources identified in RGZ, plotted as a function of the distance to the nearest optical cluster. A sharp increase in bending angle occurs near $\sim2$~Mpc, which corresponds roughly to the virial radius for typical galaxy clusters, where effects on jets from the ICM would be strongest. 
\label{fig-bendingangles}}
\end{figure*}

RGZ is assembling a large catalog of bent jets based on the automatic classifications. The bending angle is determined by selecting galaxies either with multiple discrete components or a large angular size, and then measuring the angle in the plane of the sky with respect to the host galaxy and the edges of the radio hotspots. Preliminary results reveal a large number of galaxies with sharply-bent jets, many of which are newly identified. We are analyzing the average bending angle for the sample as a function of cluster-centric distance; results show that the bending increases for galaxies closer to the center of the cluster, with a marked jump between 1--2~Mpc (Figure~\ref{fig-bendingangles}). We attribute this to the combined effects of higher peculiar velocities and an increase in the ICM density toward the cluster center (Rudnick, Willett, Garon et al., in prep). 

\section{Acknowledgments}

KWW is grateful for travel support from the American Astronomical Society and financial aid from the organizers of the EXTRA-RADSUR2015 conference. Support was also provided by NSF AST 12-11295. We thank the more than $8,000$~RGZ volunteers whose classifications make the project possible; they are individually acknowledged at \texttt{http://rgzauthors.galaxyzoo.org}.

\bibliography{kwrefs}

\end{document}